\documentclass{IEEEtaes}

\usepackage[utf8]{inputenc}
\usepackage{textcomp}
\usepackage{changes}

\usepackage{graphicx}
\usepackage{amsmath}
\usepackage{float}
\usepackage[version=4]{mhchem}
\usepackage{siunitx}
\usepackage{longtable,tabularx}
\usepackage{bm}
\usepackage{cite}
\jvol{XX}
\jnum{XX}
\jmonth{XXXXX}
\paper{1234567}
\pubyear{2022}
\doiinfo{TAES.2022.Doi Number}

\begin{document}

\title{Deep Neural Network-Based High-Precision Identification of Weak Stability Boundary Structures}

\author{Shuyue Fu}
\author{Ziqi Xu}
\author{Di Wu}
\author{Shengping Gong}
\affil{Beihang University, Beijing, People's Republic of China}

%% \author{FOURTH D. AUTHOR}
%% \affil{University of Colorado, Colorado, USA}

\receiveddate{Manuscript received XXXXX 00, 0000; revised XXXXX 00, 0000; accepted XXXXX 00, 0000.\\
This work was supported by the National Natural Science Foundation of China (Grant No. 12525204), the National Natural Science Foundation of China (Grant No. 12372044), the National Natural Science Foundation of China (Grant No. 12302058), and the Young Elite Scientists Sponsorship Program by CAST (Grant No. 2023QNRC001). }
%% \accepteddate{XXXXX XX XXXX}
%% \publisheddate{XXXXX XX XXXX}

\corresp{{\itshape (Corresponding author: Shengping Gong)}.}

\authoraddress{Shuyue Fu is with the School of Astronautics and Shen Yuan Honors College, Beihang University, Beijing, 100191, People's Republic of China (e-mail: \href{fushuyue@buaa.edu.cn}{fushuyue@buaa.edu.cn}). Ziqi Xu is with the School of Astronautics, Beihang University, Beijing, 100191, People's Republic of China (e-mail: \href{asxzq000617@163.com}{asxzq000617@163.com}). Di Wu is with the School of Astronautics, Beihang University, Beijing, 100191, People's Republic of China. He is also with the Key Laboratory of Spacecraft Design Optimization $\&$ Dynamic Simulation Technologies, Ministry of Education, Beijing, 100191, People's Republic of China (e-mail: \href{wudi2025@buaa.edu.cn}{wudi2025@buaa.edu.cn}). Shengping Gong is with the School of Astronautics, Beihang University, Beijing, 100191, People's Republic of China. He is also with the State Key Laboratory of High-Efficiency Reusable Aerospace Transportation Technology, Beijing, 102206, People's Republic of China (e-mail: \href{gongsp@buaa.edu.cn}{gongsp@buaa.edu.cn}).}

\editor{Mentions of supplemental materials and animal/human rights statements can be included here.}
\supplementary{Color versions of one or more of the figures in this article are available online at \href{http://ieeexplore.ieee.org}{http://ieeexplore.ieee.org}.}

\markboth{FU ET AL.}{DNN-BASED IDENTIFICATION METHOD OF WSB STRUCTURES}
\maketitle

\begin{abstract} Weak stability boundary structures have been widely applied to the analysis on ballistic capture and the construction of low-energy transfers. The first step of this application is to compute/identify weak stability boundary structures. Conventional numerical and analytical methods cannot simultaneously achieve computational efficiency and identification precision. In this paper, we propose an efficient and precise method to identify weak stability boundary structures based on deep neural network. The geometric and dynamical properties of weak stability boundary structures are firstly analyzed, which provides further insights into the training of the deep neural network models. Then, the optimal hyperparameter combinations are determined by examining the identification precision of the trained deep neural network models. The performance of the models with the optimal hyperparameter combinations is further validated using the representative test datasets, achieving the precision of 97.26-99.91$\%$. The trained models are also applied to constructing weak stability boundary structures.
\end{abstract}

\begin{IEEEkeywords}Planar restricted three-body problem, Weak stability boundary structure, Deep neural network
\end{IEEEkeywords}

\section{INTRODUCTION}
B{\scshape allistic} capture (BC) \cite{topputo2013optimal,li2019energy,anoe2024ballistic} is a typical phenomenon in the multi-body dynamics. Generally, it describes a scenario where the Keplerian energy with respect to the secondary body (e.g., the Moon in the Earth-Moon system) of the spacecraft is less than or equal to zero within a specific time interval (a finite time interval corresponds to a temporary BC and an infinite time interval corresponds to a permanent BC) \cite{topputo2013optimal}. When considering the natural dynamics of the multi-body model (e.g, circular restricted three-body problem (CR3BP) \cite{anoe2024ballistic} and bicircular restricted four-body problem (BCR4BP) \cite{topputo2013optimal,FU20254993}) without performing impulses, BC is usually a temporary BC \cite{topputo2013optimal}. Due to the temporary and the non-positive Keplerian energy with respect to the secondary body (usually the target body), BC takes advantage in construct transfer trajectories \cite{belbruno1993sun,topputo2015earth}, in particular low-energy transfers. Therefore, several dynamical structures have been proposed to analyze BC and consequently construct transfer trajectories, including weak stability boundary (WSB) structures \cite{belbruno1993sun,belbruno2004capture,garcia2007note,belbruno2010weak} and invariant manifolds \cite{Koon2001,moore2012trajectory,FU20254993}. These analyses have common characteristics. When using these dynamical structures to analyze BC (also construct transfers), several cuts of them under the specific ranges of the parameters (e.g., initial osculating eccentricity for WSB structures and Jacobi energy for invariant manifolds) should be obtained, which can be time-consuming in the numerical computation. To improve the efficiency to compute/identify such dynamical structures, we select one type of these structure, i.e., WSB structures as an example, and explore an efficient method to identify them.

Studies on WSB structures can be mainly categorized into three types: basic theory \cite{belbruno2004capture,garcia2007note,hyeraci2013role}, computation/identification method \cite{topputo2009computation}, and their application to the construction of low-energy transfers \cite{belbruno1993sun,ockels1999genetic,topputo2015earth}. The WSB structures can be defined as initial position sets that are able to generate multi-body trajectories with stable motion with respect to the secondary body \cite{garcia2007note,topputo2009computation,belbruno2010weak}. This concept was firstly proposed by Belbruno and Miller \cite{belbruno1993sun}, and was further refined by García and Gómez \cite{garcia2007note}. Originally, the WSB structures were computed under the fixed value of initial osculating eccentricity with respect to the secondary body \cite{garcia2007note,topputo2009computation}. Then, Fantino et al. \cite{fantino2010note} developed the computation under the fixed Jacobi energy. Such structures exist in several dynamical models, including CR3BP \cite{garcia2007note,topputo2009computation}, elliptic restricted three-body problem (ER3BP) \cite{hyeraci2010method,hyeraci2013role,dei2018survey}, and BCR4BP \cite{romagnoli2009earth}. Topputo and Belbruno \cite{topputo2009computation} developed the computation/idenfication method of WSB structures in the Sun–Jupiter planar CR3BP (PCR3BP), and several scholars have applied them to the construction of low-energy transfer trajectories \cite{belbruno1993sun,ockels1999genetic,topputo2015earth}. To construct low-energy transfers, the first step is to construct the cuts of WSB structures under several fixed values of the specific parameter (e.g., the initial osculating eccentricity). As mentioned above, this process can be time-consuming. Therefore, to further improve the computation efficiency and aid in the construction of low-energy transfers, we expect to develop an efficient method to identify WSB structures. Generally, there are two main directions to improve the computation efficiency: analytical method and method based on deep neural network (DNN). Because the precision of the analytical identification method \cite{romagnoli2009earth} should be further improved, we are devoted to developing a DNN-based identified method of WSB structures in this paper.

The DNN-based method has been widely applied to the fields of aerospace, including the classification of periodic orbits \cite{zhou2025libration}, the prediction of trajectories \cite{breen2020newton,baker2024reduced,hou2025neural}, mapping of gravity assist \cite{yan2022ann,yang2025deep}, and uncertainty propagation \cite{zhou2023neural}, and intelligent control \cite{huang2025robust,qu2025experience,wang2025robust,cheng2023neural}. It is able to capture the complex, nonlinear mapping relationship between input and output variables \cite{yang2025deep}, and mainly addresses two tasks: classification task \cite{yang2025deep} and regression task \cite{zhou2023neural}. For the identification of WSB structures, we should identify two types of samples: initial position generating trajectories with stable and unstable motion with respect to the secondary body. Therefore, the identification of WSB structures can be further transformed into a binary classification task. This exploration can be considered as a new try to establish a link between artificial intelligence technology and astrodynamics, especially the multi-body dynamics.

Based on the aforementioned discussion, the main purpose of this paper is to propose an efficient and precise method to identify WSB structures and further explore a link between artificial intelligence and multi-body dynamics. We take WSB structures in the Earth-Moon PCR3BP as an example, and construct the corresponding samples. The feature of the obtained samples is firstly analyzed. It is found that WSB structures for prograde and retrograde initial states have obvious difference. Therefore, we divide the samples into two datasets: one for prograde case and the other for retrograde case. Then, the DNN models for the binary classification task are trained, and the DNN models with the optimal hyperparameter combinations are selected. Finally, the performance of the selected models is validated using representative test datasets, with the identification precision of 97.26-99.91$\%$. The models are also applied to the construction of WSB structures under specific values of initial osculating eccentricity. The simulation results strengthen the effectiveness of the trained DNN models and proposed identification method.

The rest of this paper is organized as follows. Section \ref{sec2} presents the dynamical model and concept of WSB structures of this work. Section \ref{sec3} proposes the DNN-based method to identify WSB structures. The performance of the DNN models are validated and the selected models are applied to the construction of WSB structures in Section \ref{sec4}. Finally, conclusions are drawn in Section \ref{sec5}.

\section{WSB STRUCTURES IN THE EARTH-MOON PCR3BP}\label{sec2}
This section introduces the dynamical background of this paper, including the Earth-Moon PCR3BP, Levi-Civita regularization about the Moon, and the concepts of the weak stability boundary structures.
\subsection{Earth-Moon PCR3BP}
\begin{figure}[H]
\centerline{\includegraphics[width=15.5pc]{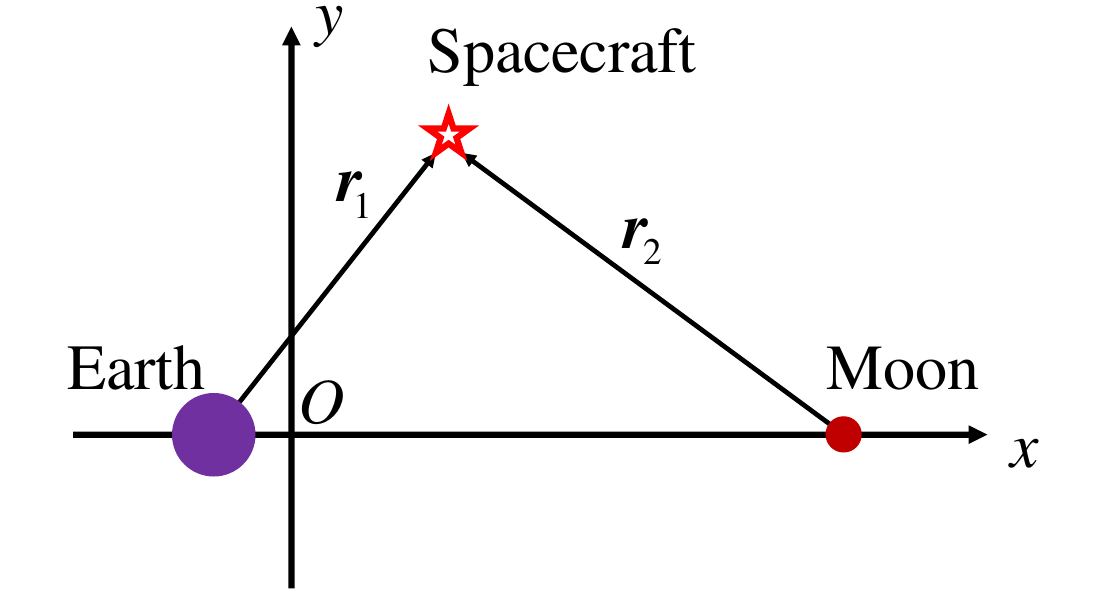}}
\caption{Earth-Moon Rotating frame.}\label{fig_EM}
\end{figure}

We focus on the weak stability boundary structures in the Earth-Moon PCR3BP. This dynamical model describes a scenario where the spacecraft moves within the Earth-Moon orbital plane. Within the Earth-Moon orbital plane, both the Earth and the Moon move in the circular orbits about their barycenter. The Earth-Moon rotating frame \cite{FU2025} is adopted to express the dynamical equations, as shown in Fig. \ref{fig_EM}. The dimensionless units are adopted as follows: the mass unit (MU) is set as the Earth-Moon combined mass, the length unit (LU) is set as the Earth-Moon distance, and the time unit is calculated by $T_\text{EM}/2\pi$, where $T_\text{EM}$ denotes the orbital period of the Earth/Moon about their barycenter. 

Then, we present the dynamical equations of the PCR3BP:
\begin{equation}
\left[ {\begin{array}{*{20}{c}}
{\begin{array}{*{20}{c}}
{\dot x}\\
{\dot y}
\end{array}}\\
{\begin{array}{*{20}{c}}
{\dot u}\\
{\dot v}
\end{array}}
\end{array}} \right] = \left[ {\begin{array}{*{20}{c}}
{\begin{array}{*{20}{c}}
u\\
v
\end{array}}\\
{\begin{array}{*{20}{c}}
{2v + \frac{{\partial {\Omega _3}}}{{\partial x}}}\\
{ - 2u + \frac{{\partial {\Omega _3}}}{{\partial y}}}
\end{array}}
\end{array}} \right]\label{eq1}
\end{equation}
\begin{equation}
{\Omega _3} = \frac{1}{2}\left[ {{x^2} + {y^2} + \mu \left( {1 - \mu } \right)} \right] + \frac{{1 - \mu }}{{{r_1}}} + \frac{\mu }{{{r_2}}}\label{eq2}
\end{equation}
\begin{equation}
{r_1} = \sqrt {{{\left( {x + \mu } \right)}^2} + {y^2}} \text{ }\text{ }\text{ }
{r_2} = \sqrt {{{\left( {x + \mu - 1} \right)}^2} + {y^2}}\label{eq3}
\end{equation}
where $\bm{X}=\left[x,\text{ }y,\text{ }u,\text{ }v\right]^{\text{T}}$ denote the states of Eq. \eqref{eq1}, $\mu$ denotes the mass parameter of the PCR3BP, ${\Omega _3}$ denotes the effective potential of the PCR3BP, $r_1$ denotes the distance between the spacecraft and the Earth, and $r_2$ denotes the distance between the spacecraft and the Moon. The specific parameters of the Earth-Moon PCR3BP can be found in Ref. \cite{FU2025}. In the PCR3BP, these exists a constant denoted as the Jacobi energy, which can be calculated by:
\begin{equation}
C = -\left( {{u^2} + {v^2}} \right) +  \left( {{x^2} + {y^2}} \right) + \frac{{2(1 - \mu) }}{{{r_1}}} + \frac{2\mu }{{{r_2}}} + \mu \left( {1 - \mu } \right) \label{eq4}
\end{equation}
To address trajectories approaching the Moon and avoid singularity, Levi-Civita regularization about the Moon is performed \cite{oshima2017analysis,oshima2019linking}:
\begin{equation}
\begin{gathered}
  x - 1 + \mu  = {u_1}^2 - {u_2}^2 \hfill \\
  y = 2{u_1}{u_2} \hfill \\
  u = \frac{{2\left( {{u_1}{u_3} - {u_2}{u_4}} \right)}}{{{u_1}^2 + {u_2}^2}} \hfill \\
  v = \frac{{2\left( {{u_2}{u_3} + {u_1}{u_4}} \right)}}{{{u_1}^2 + {u_2}^2}} \hfill \\
  {\text{d}}t = {r_2}{\text{d}}s \hfill \\ 
\end{gathered}
\label{eq5}
\end{equation}
Then, Eq. \eqref{eq1} can be transformed into:
\begin{equation}
\left[ {\begin{array}{*{20}{c}}
  {\begin{array}{*{20}{c}}
  {{u_1}^\prime } \\ 
  {{u_2}^\prime } 
\end{array}} \\ 
  {\begin{array}{*{20}{c}}
  {{u_3}^\prime } \\ 
  {{u_4}^\prime } 
\end{array}} 
\end{array}} \right] = \left[ {\begin{array}{*{20}{c}}
  {\begin{array}{*{20}{c}}
  {{u_3}} \\ 
  {{u_4}} 
\end{array}} \\ 
  {\begin{array}{*{20}{c}}
  {\frac{1}{4}\left( {a + b} \right){u_1} + \frac{1}{4}c{u_2}} \\ 
  {\frac{1}{4}\left( {a - b} \right){u_2} + \frac{1}{4}c{u_1}} 
\end{array}} 
\end{array}} \right]
\label{eq6}
\end{equation}
where $\left(\cdot\right)^\prime$ denotes $\text{d}\left(\cdot\right)/\text{d}s$, and $a$, $b$, and $c$ are expressed as:
\begin{align}\label{eq7}
a &= \frac{{2\left( {1 - \mu } \right)}}{{\sqrt {{{\left( {{u_1}^2 - {u_2}^2 + 1} \right)}^2} + 4{u_1}^2{u_2}^2} }} - C \\ 
\notag&+ {\left( {{u_1}^2 - {u_2}^2 + 1 - \mu } \right)^2} + 4{u_1}^2{u_2}^2 + \mu \left( {1 - \mu } \right)
\end{align}
\begin{align}\label{eq8}
b &= 8\left( {{u_2}{u_3} + {u_1}{u_4}} \right) \\
\notag&+ 2\left( {{u_1}^2 + {u_2}^2} \right)\left( {{u_1}^2 - {u_2}^2 + 1 - \mu } \right) \\
\notag&- \frac{{2\left( {1 - \mu } \right)\left( {{u_1}^2 - {u_2}^2 + 1} \right)\left( {{u_1}^2 + {u_2}^2} \right)}}{{{{\left( {\sqrt {{{\left( {{u_1}^2 - {u_2}^2 + 1} \right)}^2} + 4{u_1}^2{u_2}^2} } \right)}^3}}}
\end{align}
\begin{align}\label{eq9}
c &= 4{u_1}{u_2}\left( {{u_1}^2 + {u_2}^2} \right) - 8\left( {{u_1}{u_3} - {u_2}{u_4}} \right) \\
\notag & - \frac{{4\left( {1 - \mu } \right){u_1}{u_2}\left( {{u_1}^2 + {u_2}^2} \right)}}{{{{\left( {\sqrt {{{\left( {{u_1}^2 - {u_2}^2 + 1} \right)}^2} + 4{u_1}^2{u_2}^2} } \right)}^3}}}
\end{align}
With $\left[u_1,\text{ }u_2,\text{ }u_3,\text{ }u_4\right]^\text{T}$, the Jacobi energy $C$ is expressed as:
\begin{align}\label{eq10}
C &=  - \frac{{4\left( {{u_3}^2 + {u_4}^2} \right)}}{{{u_1}^2 + {u_2}^2}} + \frac{{2\mu }}{{{u_1}^2 + {u_2}^2}} + \mu \left( {1 - \mu } \right)\\
\notag&+ \left[ {{{\left( {{u_1}^2 - {u_2}^2 + 1 - \mu } \right)}^2} + 4{u_1}^2{u_2}^2} \right]  \\
\notag&+ \frac{{2\left( {1 - \mu } \right)}}{{\sqrt {{{\left( {{u_1}^2 - {u_2}^2 + 1} \right)}^2} + 4{u_1}^2{u_2}^2} }} 
\end{align}
Then, we adopt Eq. \eqref{eq6} to integrate the three-body trajectories and obtain the weak stability boundary (WSB) structures, using the variable step-size, variable order (VSVO) Adams-Bashforth-Moulton algorithm with absolute and relative tolerances set to $1 \times 10^{-13}$ \cite{oshima2021capture}. Subsequently, the concept of the WSB structures is introduced.

\subsection{WSB Structures}\label{subsection2.2}
\begin{figure}[H]
\centerline{\includegraphics[width=17.5pc]{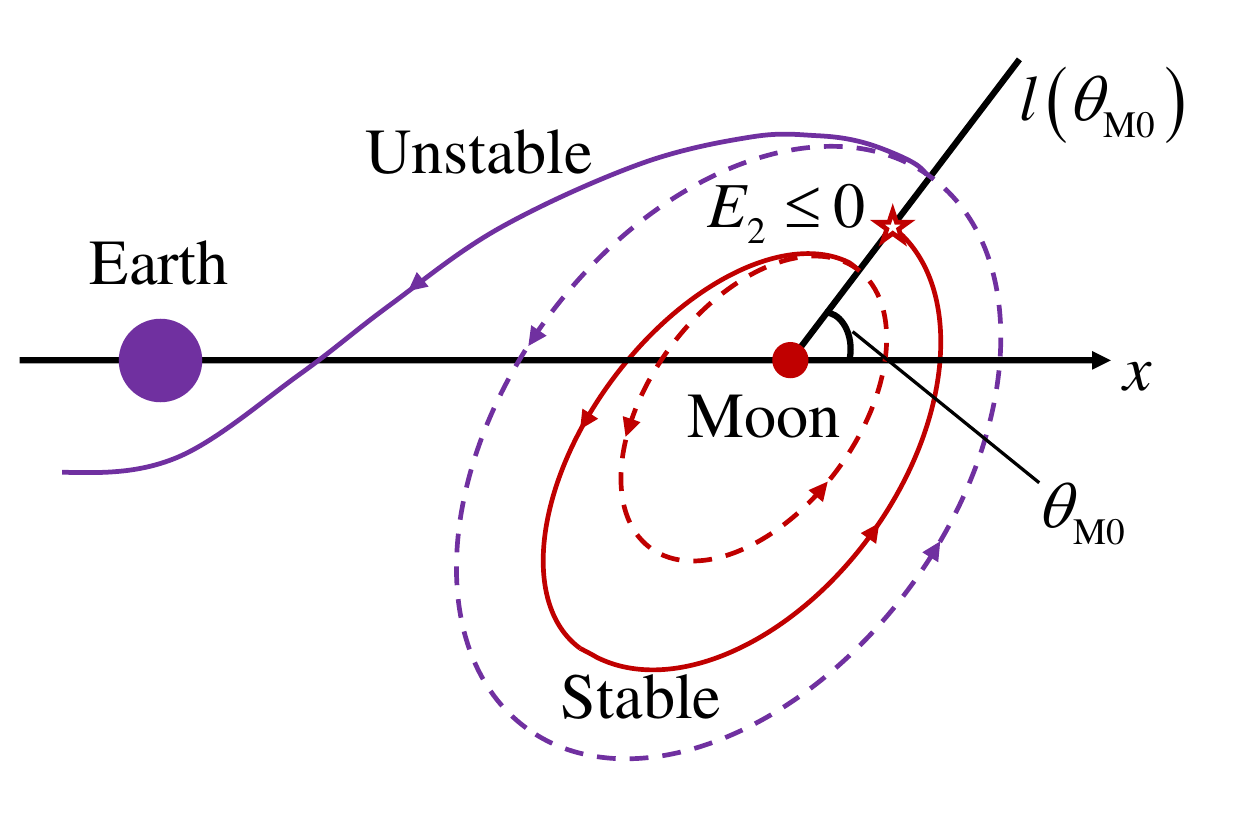}}
\caption{Schematic of stable and unstable trajectories.}\label{fig_WSB}
\end{figure}

In this subsection, we present the concept of the WSB structures in the Earth-Moon PCR3BP. This concept was firstly proposed by Belbruno and Miller \cite{belbruno1993sun} and refined by García and Gómez \cite{garcia2007note}. The WSB structures have been widely used in the analysis on the ballistic capture \cite{hyeraci2010method} and the construction of low-energy transfer trajectories \cite{belbruno1993sun}. Here, following Refs. \cite{garcia2007note,belbruno2010weak}, we present the definition of the stable motion around the Moon. As shown in Fig. \ref{fig_WSB}, the initial states $\bm{X}_0$ of the considered trajectories are selected on $l\left(\theta_\text{M0}\right)$ ($\theta_\text{M0}$ denotes the initial Moon phase angle calculated by $\text{atan2}\left(y_0,\text{ }x_0+\mu-1\right)$) and the considered trajectories depart $l\left(\theta_\text{M0}\right)$ with tangential velocities. The subscript “0” denotes the quantities associated with the initial epoch. 

The initial states satisfy $E_{20}<0$, where $E_2$ denotes the Keplerian energy with respect to the Moon and is expressed as:
\begin{equation}
{E_{2}} = \frac{1}{2}\left[ {{{\left( {{u} - {y}} \right)}^2} + {{\left( {{v} + {x} + \mu  - 1} \right)}^2}} \right] - \frac{\mu }{r_{2}} \label{eq11}
\end{equation}
Considering two cases: prograde and retrograde motion \cite{garcia2007note,FU20254993}, two sets of initial states can be obtained as:
\begin{equation}
\begin{gathered}
 {\text{Prograde:}} \hfill \\
  {x_0} = {r_{20}}\cos {\theta _{\text{M0}}} + 1 - \mu  \hfill \\
  {y_0} = {r_{20}}\sin {\theta _{\text{M0}}} \hfill \\
  {u_0} =  - \left( {\sqrt {\frac{{\mu \left( {1 + e} \right)}}{{{r_{20}}}}}  - {r_{20}}} \right)\sin {\theta _{\text{M0}}} \hfill \\
  {v_0} = \left( {\sqrt {\frac{{\mu \left( {1 + e} \right)}}{{{r_{20}}}}}  - {r_{20}}} \right)\cos {\theta _{\text{M0}}} \hfill \\ 
\end{gathered} 
 \label{eq12}
\end{equation}

\begin{equation}
\begin{gathered}
  {\text{Retrograde:}} \hfill \\
  {x_0} = {r_{20}}\cos {\theta _{\text{M0}}} + 1 - \mu  \hfill \\
  {y_0} = {r_{20}}\sin {\theta _{\text{M0}}} \hfill \\
  {u_0} = \left( {\sqrt {\frac{{\mu \left( {1 + e} \right)}}{{{r_{20}}}}}  + {r_{20}}} \right)\sin {\theta _{\text{M0}}} \hfill \\
  {v_0} =  - \left( {\sqrt {\frac{{\mu \left( {1 + e} \right)}}{{{r_{20}}}}}  + {r_{20}}} \right)\cos {\theta _{\text{M0}}} \hfill \\ 
\end{gathered}
 \label{eq13}
\end{equation}
where $r_{20}$ denotes the initial distance between the spacecraft and the Moon on $l\left(\theta_\text{M0}\right)$, and $e$ denotes the initial osculating eccentricity with respect to the Moon at the initial epoch. The velocity $\sqrt {\frac{{\mu \left( {1 + e} \right)}}{{{r_{20}}}}}$ denotes the initial velocity of the spacecraft in the Moon-centered inertial frame. With these states, $E_{20}$ can be expressed as:
\begin{equation}
{E_{20}} = \frac{1}{2}\frac{\mu\left(e-1\right) }{r_{20}} \label{eq14}
\end{equation}
Therefore, the condition that $E_{20}<0$ is equivalent to the condition that $e<1$. When initial states are obtained from Eq. \eqref{eq12} or Eq. \eqref{eq13}, these states are transformed into $\left[u_1,\text{ }u_2,\text{ }u_3,\text{ }u_4\right]^\text{T}$. Trajectories are integrated numerically. The stable motion is identified by the following definition:
\begin{itemize}
\item The motion is considered as the stable motion if the value of $E_2$ satisfies $E_2 \leq 0$ when the trajectory completes a revolution around the Moon without going around the Earth. Otherwise, the motion is considered unstable.
\end{itemize}
Numerically, the integration time is set to $T = \int_0^{5000} {{r_2}{\text{d}}s}$. When the states of trajectories satisfy $x<-\mu$ and $y=0$, the integration terminates. We consider that the trajectory completes a revolution around the Moon when the states are able to satisfy \cite{topputo2009computation}:
\begin{equation}
\left|\theta_\text{M}-\theta_\text{M0}\right|=2\pi \label{eq15}
\end{equation}
Following the aforementioned definition of the stable motion, the WSB structures $\mathcal{W}\left(e\right)$ denote the $\left(x_0,\text{ }y_0\right)$ set whose corresponding trajectories are identified as the stable motion under the fixed value of $e$:
\begin{equation}
\mathcal{W}\left(e\right)=\{\left(x_0,\text{ }y_0\right)|E_2\leq 0 \text{ }\text{when}\text{ }\left|\theta_\text{M}-\theta_\text{M0}\right|=2\pi\}
\label{eq15_new}
\end{equation}
According the types of initial states, the WSB structures are categorized into the prograde and retrograde WSB structures. When the WSB structures are applied to the construction of ballistic capture and low-energy transfer trajectories, the first step to obtain several cuts of the WSB structures under the fixed  value of $e$ \cite{garcia2007note}. This process can be time-consuming. To improve the computational efficiency and further aid in the design of ballistic capture and low-energy transfer trajectories, we propose an identification method of WSB structures using the DNN, which is detailed in Section \ref{sec3}.

\section{DNN-BASED IDENTIFICATION METHOD OF WSB STRUCTURES}\label{sec3}
In this section, we propose the identification method of WSB structures using the DNN, including constructing samples, analyzing the features of the samples, and training and evaluating DNN model. Here, the identification of WSB structures is transformed into a binary classification task, and detailed procedure can be found in the following texts. 

\subsection{Constructing Samples and Analyzing Features}\label{subsection3.1}
\begin{figure*}
\centerline{\includegraphics[width=37pc]{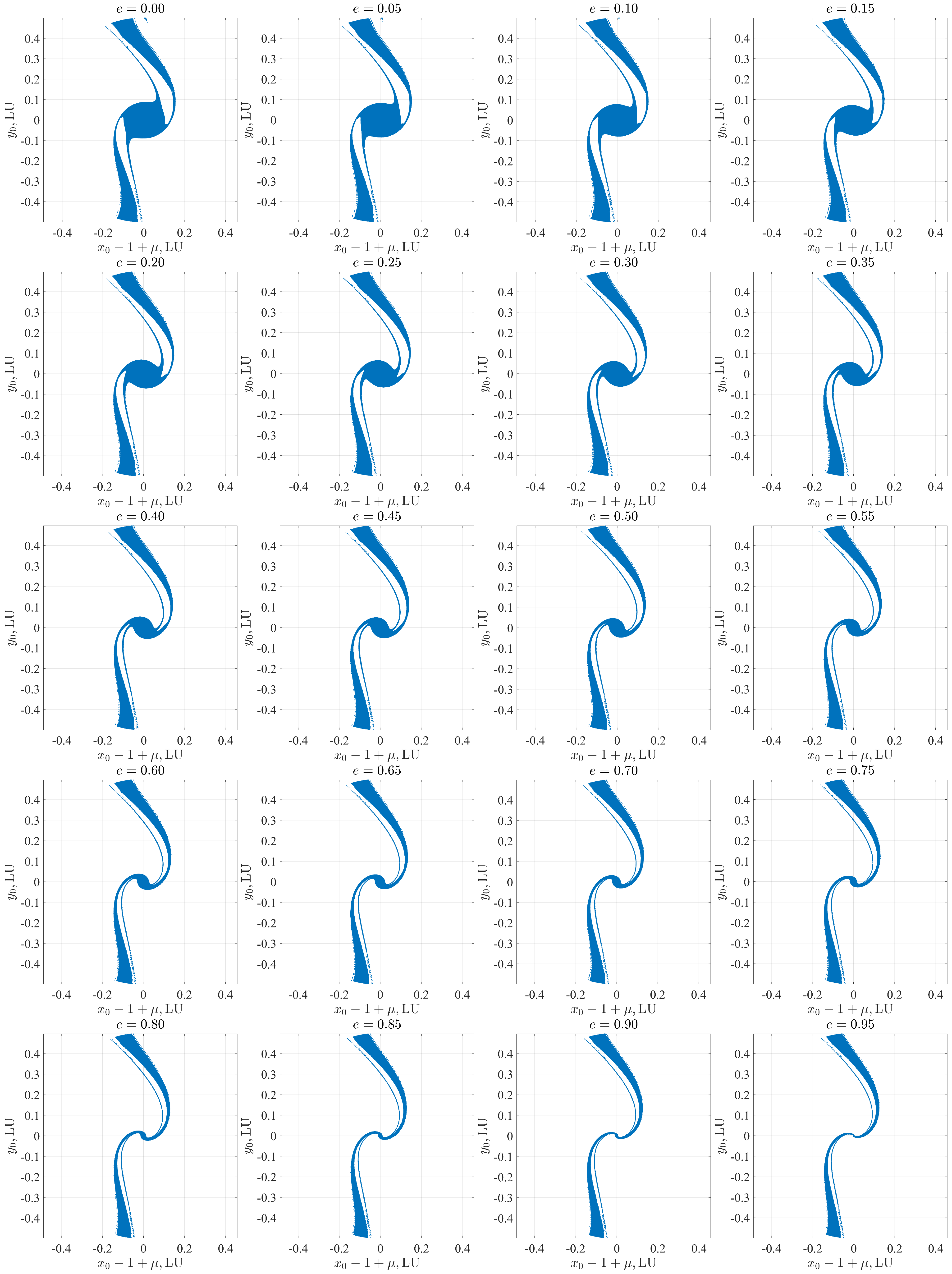}}
\caption{Generated WSB structures (prograde case).}\label{fig_p_WSB}
\end{figure*}

\begin{figure*}
\centerline{\includegraphics[width=37pc]{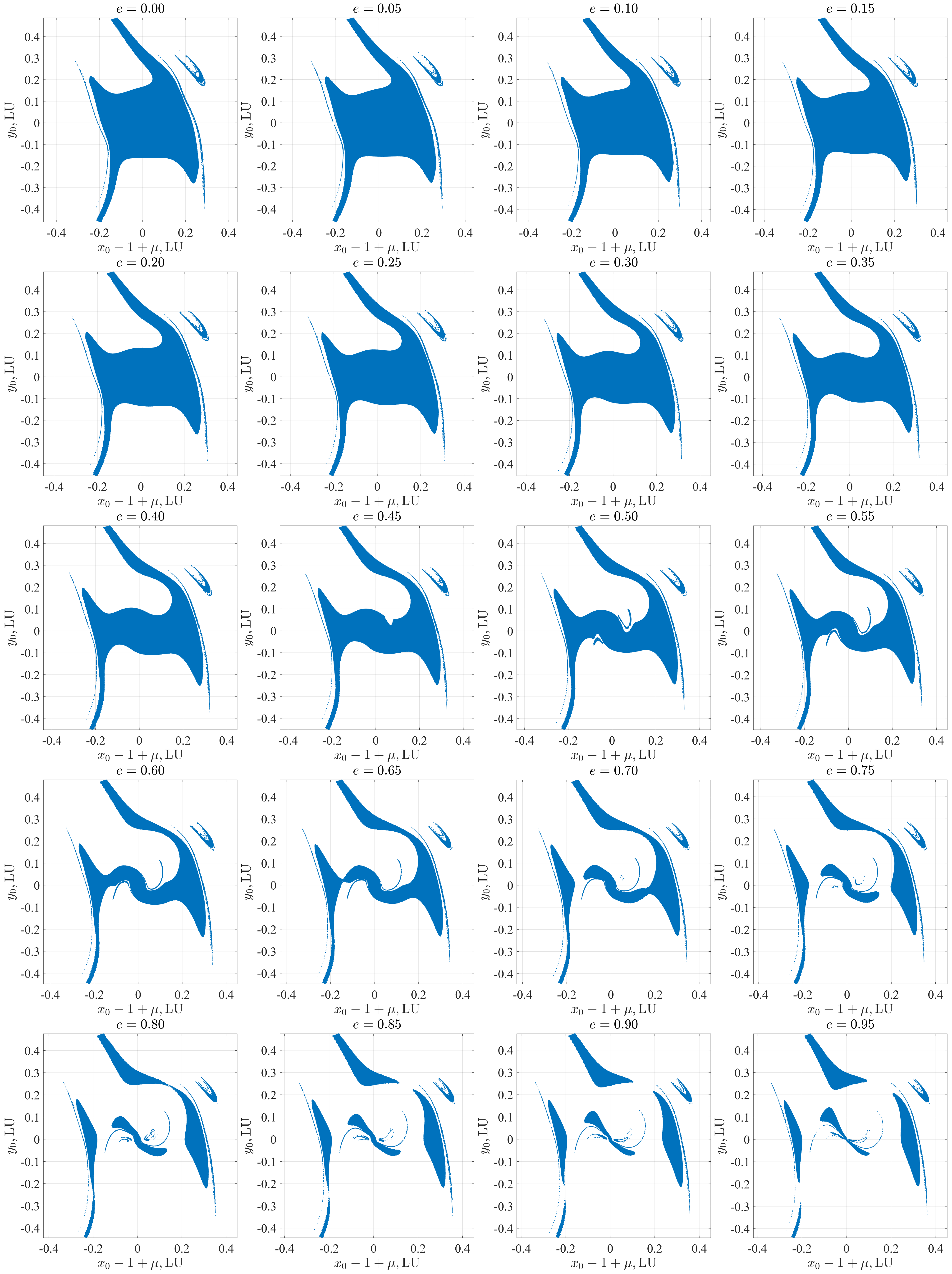}}
\caption{Generated WSB structures (retrograde case).}\label{fig_r_WSB}
\end{figure*}

\begin{figure*}
\centerline{\includegraphics[width=33pc]{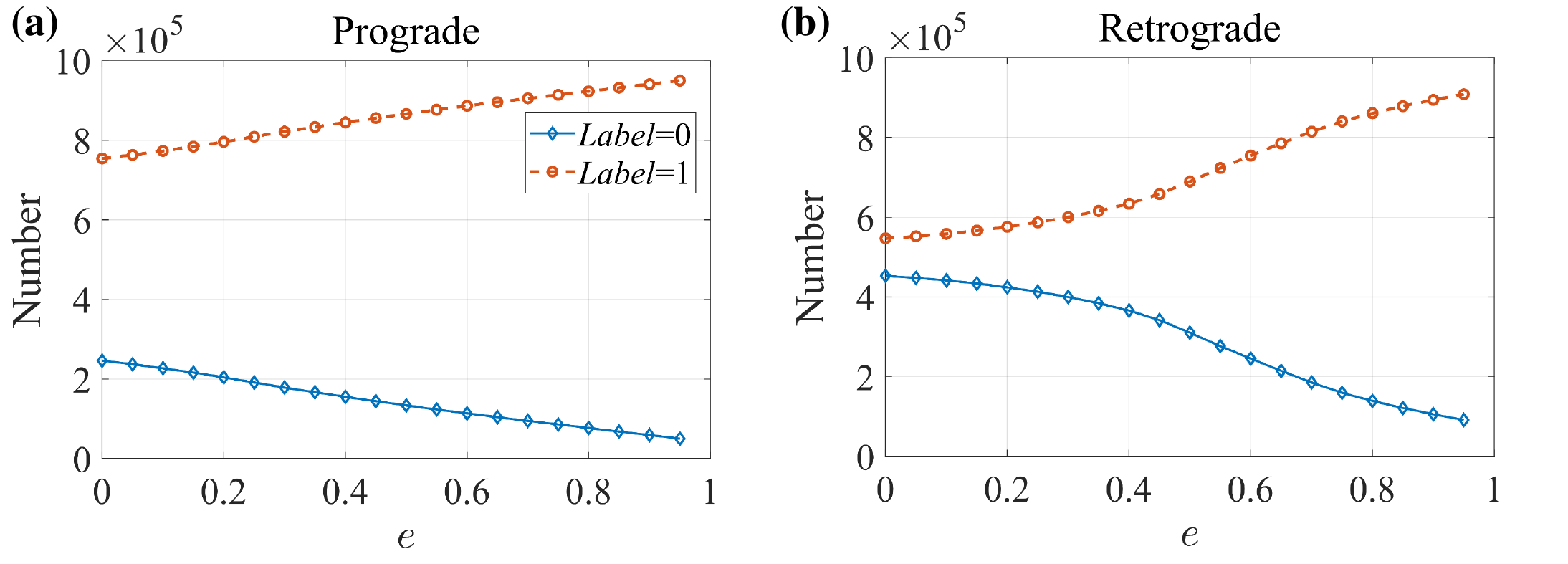}}
\caption{Number of points with $Label=0$ and $Label=1$. (a) Prograde case; (b) Retrograde case.}\label{fig_pr}
\end{figure*}

Based on the discussion in Section II-\ref{subsection2.2}, when $e$, $r_{20}$, and $\theta_\text{M0}$ are given, we can identify whether the motion of the trajectory integrated by the initial states in Eq. \eqref{eq12} or Eq. \eqref{eq13} is stable. Therefore, the input vector can be set as:
\begin{equation}
\bm{s}_0=\left[e,\text{ }r_{20},\text{ }\theta_\text{M0}\right]^\text{T} \label{eq16}
\end{equation}
and the output of the samples is the labels (denoted as $Label$) 0 and 1 (0 for the stable motion and 1 for the unstable motion). We set $e \in \left[0,\text{ }1\right)$ with a step-size of 0.05, $r_{20} \in \left(0,\text{ }0.5\right]\text{ }\left(\text{LU}\right)$ with a step-size of 0.0005 LU, and $\theta_\text{M0} \in \left[0,\text{ }2\pi\right)$ with a step-size of $\pi/500$ to generate the samples. To further improve the performance of the trained model, we transform $\bm{s}_0$ into the following form:
\begin{equation}
\bm{s}=\left[e,\text{ }\log_{10}\left(r_{20}+\epsilon\right),\text{ }\cos \theta_\text{M0},\text{ }\sin \theta_\text{M0}\right]^\text{T} \label{eq17}
\end{equation}
where $\epsilon$ is set to $1\times 10^{-12}$. The generated samples are shown in Figs. \ref{fig_p_WSB}-\ref{fig_r_WSB}. From these figures, different geometric configurations between prograde and retrograde WSB structures are observed, implying different dynamical properties between prograde and retrograde cases. For a single case, the configurations of WSB structures are varying with the value of $e$. As the value of $e$ increases, the number of samples with $Label=0$ (i.e., stable points) decreases, while the number of samples with $Label=1$ increases, as shown in Fig. \ref{fig_pr}. Generally, an imbalanced distribution of samples with $Label=0$ and $Label=1$ can be observed for both prograde and retrograde cases. Based on the aforementioned discussion, we divide the generated samples into two datasets: one for progarde case, and the other for retrograde case, denoted as Datasets 1 and 2, respectively. The imbalanced distribution of samples with $Label=0$ and $Label=1$ is also used to provide further insight into the DNN training and evaluation, which is detailed in the next subsection.

\subsection{Training and Evaluating DNN Models}\label{subsection3.2}
In this subsection, we present the training settings of the DNN models. The DNN models are trained using PyTorch 2.9.0 (CUDA 13.0) in a Python 3.12.12 environment. We set the random seed to 42 to ensure reproducibility. Datasets 1 and 2 are respectively divided into the training dataset (80$\%$) and the validation dataset (20$\%$). The validation dataset is used to evaluate the trained models preliminarily and determine the hyperparameter combinations used in the training. When training the DNN models, we normalize the transformed input vector $\bm{s}$ to $\left[-1,\text{ }1\right]$ using \textit{MinMaxScaler} function from sklearn. The DNN model discussed in this paper is presented in Fig. \ref{fig_DNN} (a). There are three types of layers in the considered model: input layer, hidden layers, and output layer. Figure \ref{fig_DNN} (b) presents the schematic of neuron computation in DNN. In each neuron computation, the mapping from the input $\left[x_1,\text{ }x_2,\text{ }...,\text{ }x_n\right]^\text{T}$ to the output $h$ can be expressed as \cite{yang2025deep}:
\begin{figure}
\centerline{\includegraphics[width=17.5pc]{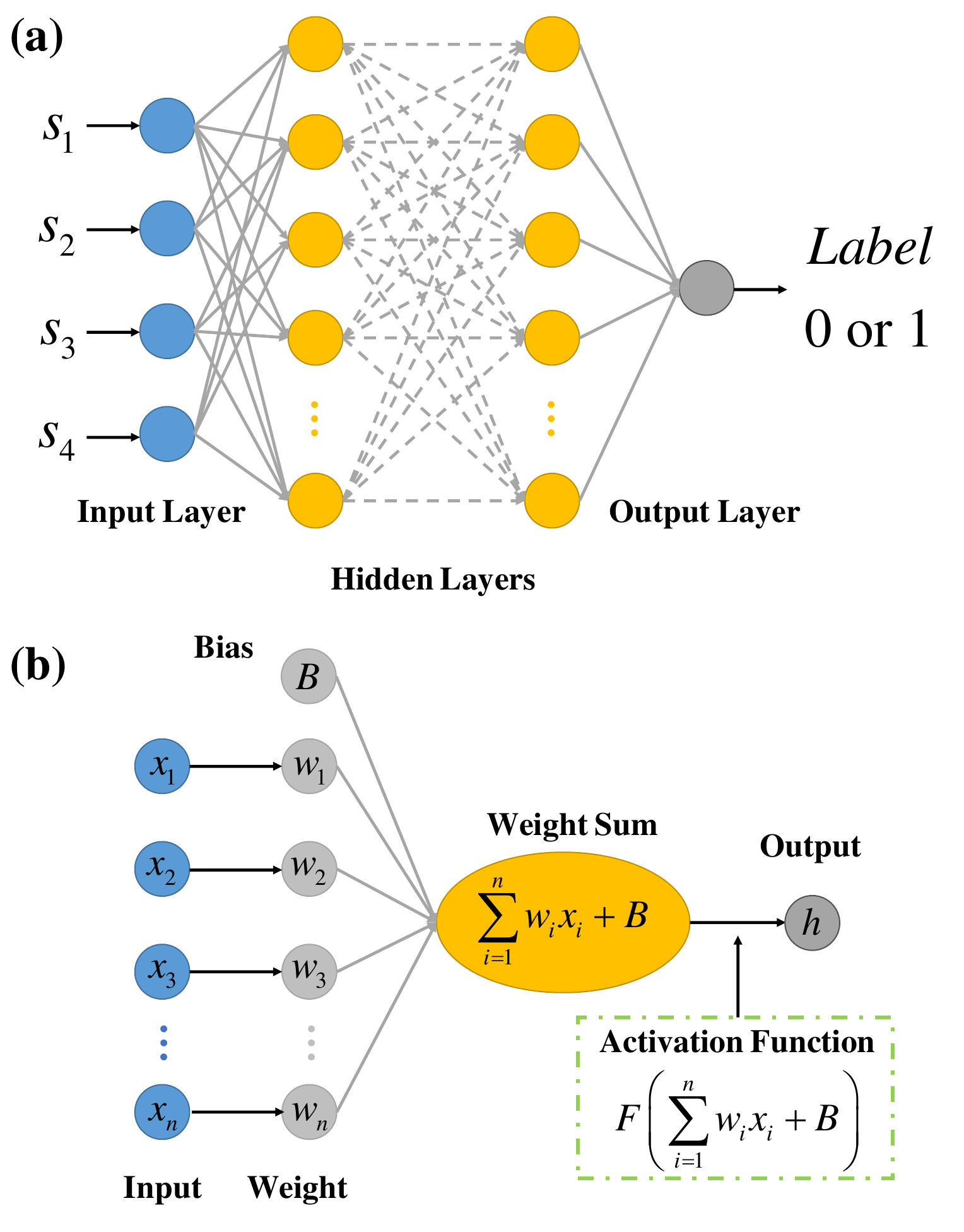}}
\caption{The structure of the discussed DNN model and the schematic of neuron computation in DNN. (a) DNN structure; (b) Neuron computation in DNN.}\label{fig_DNN}
\end{figure}
\begin{equation}
h = F\left( {\sum\limits_{i = 1}^n {{w_i}{x_i} + {B}} } \right) \label{eq18}
\end{equation}
where $F\left(\cdot\right)$ denotes the activation function, $w_i\text{ }\left(i=1,\text{ }...,\text{ }n\right)$ denotes the weights of the corresponding input variables, and $B$ denotes the bias. In the considered DNN model, the activation function is fixed as \textit{Tanh} except for the output layer. For the binary classification task constructed in this paper, there is no activation function when training the DNN models because the loss function is set as \textit{BCEWithLogitsLoss}. When evaluating and applying the trained model, the activation function \textit{Sigmoid} is used to transform the scalar output into a probability \cite{subramanian2018deep,godoy2022deep}. This probability is further transformed into the classification labels through the specific threshold (discussed in the following texts). Then, the settings of the considered DNN models are summarized as follows:
\begin{itemize}
\item \textit{Structure of DNN models:} The number of hidden layers is set as 3, and the number of neurons per hidden layer ($n_\text{neurons}$) is treated as the hyperparameter to be determined. The $n_\text{neurons}$ combined with hidden layer is denoted as $Layersize$, and the corresponding settings are presented in Table \ref{tab2}.
\item \textit{Training:} During the training, the \textit{Adam} optimizer because of its remarkable computational efficiency and convergence performance \cite{zhou2023neural}. The loss function is set as \textit{BCEWithLogitsLoss}, and the learning ratio ($lr$) is treated as the hyperparameter (see Table \ref{tab2}). The ratio of weight decay to $lr$ is fixed as 0.001. The training epoch is set as 500, and the batch size ($Batchsize$) is treated as the hyperparameter.
\end{itemize}

\begin{table}[h]
\centering
\caption{Settings of Hyperparameters}\label{tab2}%
\begin{tabular}{@{}ll@{}}
\hline
Hyperparameter & Setting   \\
\hline
$Layersize$    & $\left[64,\text{ }64,\text{ }64\right]$ \\
    & $\left[64,\text{ }64,\text{ }32\right]$ \\
        & $\left[64,\text{ }32,\text{ }32\right]$ \\
     & $\left[64,\text{ }32,\text{ }16\right]$ \\
$lr$ & $0.1$, $0.01$, $0.001$ \\
$Batchsize$ & $51200$, $102400$ \\
\hline
\end{tabular}
\end{table}
The validation dataset is used to used to evaluate the trained models preliminarily and determine the optimal hyperparameter combinations. There are several metrics to evaluate the performance of the trained models, such as accuracy, precision, and recall \cite{subramanian2018deep}. In this paper, we consider the samples with stable motion ($Label=0$) as the positive samples, and consequently the samples with unstable motion ($Label=1$) as the negative samples. Then, four numbers are introduce: $TP$ (the number of correctly identified positive samples), $TN$ (the number of correctly identified negative samples), $FP$ (the number of incorrectly identified positive samples), and $FN$ (the number of incorrectly identified negative samples). In this paper, the purpose of constructing DNN models to identify WSB structures is to improve computational efficiency and aid in further construction of multi-body trajectories based on the WSB structures. Because the obtained datasets are imbalanced datasets, only one type of sample (positive or negative) is focused on \cite{godoy2022deep}. Because we mainly use the $\left(x_0,\text{ }y_0\right)$ belonging to the WSB structures in the construction of ballistic capture and low-energy transfer trajectories, the positive samples are focused on. Moreover, the construction further requires that the identified points of the WSB structures is precise as possible. The number $TP$ is especially focused on. Therefore, we adopt the precision ($Precision$) as the metric to evaluate the performance of the trained DNN models:
\begin{equation}
Precision = \frac{TP}{TP+FP} \label{eq19}
\end{equation}
The threshold of probability is set to the value less than 0.5 to further improve the precision of the models \cite{godoy2022deep}. In this paper, this value is specified as 0.4, i.e., when the probability is larger than 0.4, it is transformed into $Label=1$. By calculating the precision of each trained model, we select the final models with the optimal hyperparameter combinations whose precision are highest. Then, to further evaluate the performance of the selected models, we generate other datasets as the test dataset. We set $e =0.03,\text{ }0.52,\text{ }0.93$, $r_{20} \in \left(0,\text{ }0.5\right]\text{ }\left(\text{LU}\right)$ with a step-size of 0.0005 LU, and $\theta_\text{M0} \in \left[0,\text{ }2\pi\right)$ with a step-size of $\pi/500$ to generate the datasets. We construct the dataset under the same value of $e$ for these two cases (prograde and retrograde cases), i.e., 6 test datasets are constructed in total. It is worth to note that this further validation can also be considered as the application of the trained DNN models to the construction of the WSB structure, which is the first step of the WSB-based construction of ballistic capture and low-energy transfer trajectories. Then, the results are presented and discussed in the next section.

\textit{Remark:} In this paper, we adopt the Earth-Moon PCR3BP to generate the WSB structures and propose the DNN-based method to identify them. The proposed method can be further extended to the identification of the WSB structures in other PCR3BP (or planar ER3BP) systems with different values of $\mu$ \cite{topputo2009computation,hyeraci2010method,hyeraci2013role,dei2018survey,fu2025analytical} and the Earth-Moon PCR3BP with perturbations \cite{romagnoli2009earth,fu2025four,yin2023midcourse,wang2025mechanism}. To achieve this extend, the datasets and samples of the WSB structures in the different dynamical models should be constructed.

\section{RESULTS AND DISCUSSION}\label{sec4}
In this section, we first select the DNN models for progarde and retrograde WSB structures based on the identification precision. Then, the performance of the selected models are further validated by the test datasets (also be considered that the selected models are applied to the construction of WSB structures under different values of $e$).
\subsection{Selection of DNN Models}\label{subsection4.1}
Following the method proposed in Section \ref{sec3}, we trained 48 DNN models (24 for the prograde case and 24 for the retrograde case) with the hyperparameters presented in Table \ref{tab2}. Then, we select the DNN models with the optimal hyperparameter combinations (i.e., the models with the highest precision for prograde and retrograde cases). As presented in Tables \ref{tab3}-\ref{tab4}, the identification precision increases as the decrease of $lr$. Then, we present the optimal hyperparameter combinations for these two cases as follows:

\begin{itemize}
\item \textit{Progarde case:} $Layersize=\left[64,\text{ }64,\text{ }64\right]$, $lr=0.001$, $Batchsize=51200$, the identification precision is 99.31$\%$;
\item \textit{Retrogarde case:} $Layersize=\left[64,\text{ }32,\text{ }32\right]$, $lr=0.001$, $Batchsize=102400$, the identification precision is 99.80$\%$.
\end{itemize}

\begin{table}[h]
\centering
\caption{Precision under Different Hyperparameter Combinations (Progarde Case)}\label{tab3}%
\begin{tabular}{@{}llll@{}}
\hline
$Layersize$ & $lr$ & $Batchsize$ & $Precision$, $\%$   \\
\hline
$\left[64,\text{ }64,\text{ }64\right]$ & $0.1$ & $51200$ & $83.16$ \\
$\left[64,\text{ }64,\text{ }32\right]$ & $0.1$ & $51200$ & $93.69$ \\
$\left[64,\text{ }32,\text{ }32\right]$ & $0.1$ & $51200$ & $97.83$ \\
$\left[64,\text{ }32,\text{ }16\right]$ & $0.1$ & $51200$ & $96.63$ \\
$\left[64,\text{ }64,\text{ }64\right]$ & $0.1$ & $102400$ & $95.99$ \\
$\left[64,\text{ }64,\text{ }32\right]$ & $0.1$ & $102400$ & $96.31$ \\
$\left[64,\text{ }32,\text{ }32\right]$ & $0.1$ & $102400$ & $96.43$ \\
$\left[64,\text{ }32,\text{ }16\right]$ & $0.1$ & $102400$ & $95.38$ \\
$\left[64,\text{ }64,\text{ }64\right]$ & $0.01$ & $51200$ & $98.80$ \\
$\left[64,\text{ }64,\text{ }32\right]$ & $0.01$ & $51200$ & $98.79$ \\
$\left[64,\text{ }32,\text{ }32\right]$ & $0.01$ & $51200$ & $98.87$ \\
$\left[64,\text{ }32,\text{ }16\right]$ & $0.01$ & $51200$ & $98.62$ \\
$\left[64,\text{ }64,\text{ }64\right]$ & $0.01$ & $102400$ & $98.60$ \\
$\left[64,\text{ }64,\text{ }32\right]$ & $0.01$ & $102400$ & $98.60$ \\
$\left[64,\text{ }32,\text{ }32\right]$ & $0.01$ & $102400$ & $98.87$ \\
$\left[64,\text{ }32,\text{ }16\right]$ & $0.01$ & $102400$ & $98.80$ \\
$\left[64,\text{ }64,\text{ }64\right]$ & $0.001$ & $51200$ & $99.31$ \\
$\left[64,\text{ }64,\text{ }32\right]$ & $0.001$ & $51200$ & $99.17$ \\
$\left[64,\text{ }32,\text{ }32\right]$ & $0.001$ & $51200$ & $99.17$ \\
$\left[64,\text{ }32,\text{ }16\right]$ & $0.001$ & $51200$ & $99.20$ \\
$\left[64,\text{ }64,\text{ }64\right]$ & $0.001$ & $102400$ & $99.20$ \\
$\left[64,\text{ }64,\text{ }32\right]$ & $0.001$ & $102400$ & $99.17$ \\
$\left[64,\text{ }32,\text{ }32\right]$ & $0.001$ & $102400$ & $99.24$ \\
$\left[64,\text{ }32,\text{ }16\right]$ & $0.001$ & $102400$ & $99.01$ \\
\hline
\end{tabular}
\end{table}

\begin{table}[h]
\centering
\caption{Precision under Different Hyperparameter Combinations (Retrogarde Case)}\label{tab4}%
\begin{tabular}{@{}llll@{}}
\hline
$Layersize$ & $lr$ & $Batchsize$ & $Precision$, $\%$   \\
\hline
$\left[64,\text{ }64,\text{ }64\right]$ & $0.1$ & $51200$ & $80.27$ \\
$\left[64,\text{ }64,\text{ }32\right]$ & $0.1$ & $51200$ & $97.98$ \\
$\left[64,\text{ }32,\text{ }32\right]$ & $0.1$ & $51200$ & $98.63$ \\
$\left[64,\text{ }32,\text{ }16\right]$ & $0.1$ & $51200$ & $96.30$ \\
$\left[64,\text{ }64,\text{ }64\right]$ & $0.1$ & $102400$ & $97.44$ \\
$\left[64,\text{ }64,\text{ }32\right]$ & $0.1$ & $102400$ & $98.99$ \\
$\left[64,\text{ }32,\text{ }32\right]$ & $0.1$ & $102400$ & $98.24$ \\
$\left[64,\text{ }32,\text{ }16\right]$ & $0.1$ & $102400$ & $96.45$ \\
$\left[64,\text{ }64,\text{ }64\right]$ & $0.01$ & $51200$ & $99.40$ \\
$\left[64,\text{ }64,\text{ }32\right]$ & $0.01$ & $51200$ & $99.70$ \\
$\left[64,\text{ }32,\text{ }32\right]$ & $0.01$ & $51200$ & $99.67$ \\
$\left[64,\text{ }32,\text{ }16\right]$ & $0.01$ & $51200$ & $99.43$ \\
$\left[64,\text{ }64,\text{ }64\right]$ & $0.01$ & $102400$ & $99.38$ \\
$\left[64,\text{ }64,\text{ }32\right]$ & $0.01$ & $102400$ & $99.74$ \\
$\left[64,\text{ }32,\text{ }32\right]$ & $0.01$ & $102400$ & $99.63$ \\
$\left[64,\text{ }32,\text{ }16\right]$ & $0.01$ & $102400$ & $99.40$ \\
$\left[64,\text{ }64,\text{ }64\right]$ & $0.001$ & $51200$ & $99.64$ \\
$\left[64,\text{ }64,\text{ }32\right]$ & $0.001$ & $51200$ & $99.70$ \\
$\left[64,\text{ }32,\text{ }32\right]$ & $0.001$ & $51200$ & $99.72$ \\
$\left[64,\text{ }32,\text{ }16\right]$ & $0.001$ & $51200$ & $99.60$ \\
$\left[64,\text{ }64,\text{ }64\right]$ & $0.001$ & $102400$ & $99.74$ \\
$\left[64,\text{ }64,\text{ }32\right]$ & $0.001$ & $102400$ & $99.69$ \\
$\left[64,\text{ }32,\text{ }32\right]$ & $0.001$ & $102400$ & $99.80$ \\
$\left[64,\text{ }32,\text{ }16\right]$ & $0.001$ & $102400$ & $99.60$ \\
\hline
\end{tabular}
\end{table}

\subsection{Validation of Selected DNN Models Using Test Datasets and Application}\label{subsection4.2}
\begin{figure*}
\centerline{\includegraphics[width=33pc]{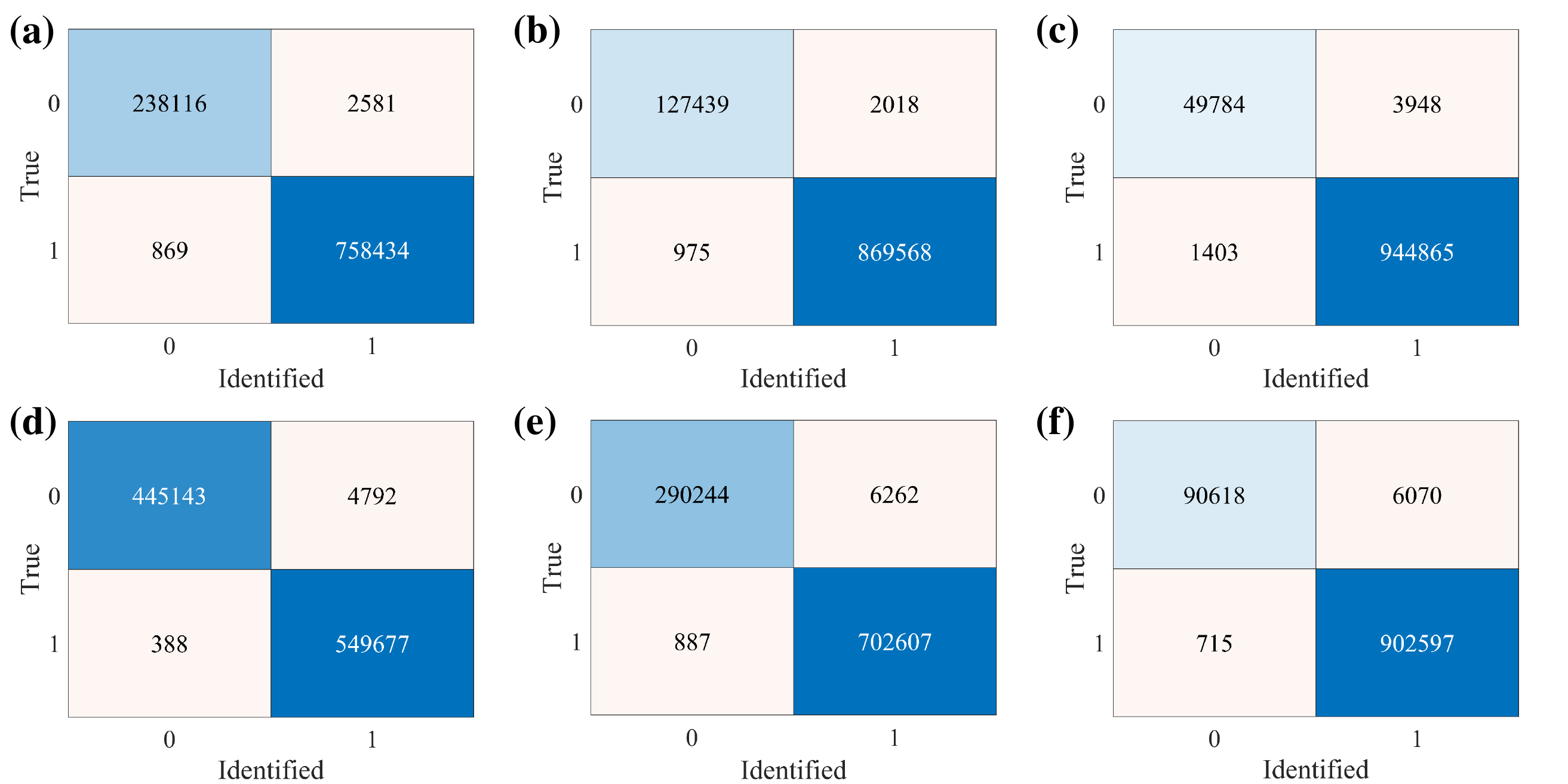}}
\caption{Confusion matrices of identification. (a) Prograde, $e=0.03$; (b) Prograde, $e=0.52$; (c) Prograde, $e=0.93$; (d) Retrograde, $e=0.03$; (e) Retrograde, $e=0.52$; (f) Retrograde, $e=0.93$.}\label{fig_CM}
\end{figure*}

After selecting the DNN models, we further validate their performance using the six test datasets mentioned in Section III-\ref{subsection3.2}. The corresponding identification precision is presented in Table \ref{tab5}. From Table \ref{tab5}, it can be observed that the precision of the selected DNN models remain high (97.26-99.91$\%$) using the six test datasets. In particular, as the value of $e$ increases, the precision decreases both for prograde and retrograde cases. The corresponding confusion matrices of the identification are presented in Fig. \ref{fig_CM}.

\begin{table}[h]
\centering
\caption{Identification Precision of Six Test Datasets}\label{tab5}%
\begin{tabular}{@{}lll@{}}
\hline
Case & $e$  & $Precision$, $\%$   \\
\hline
Prograde & $0.03$ & $99.64$ \\
 & $0.52$ & $99.24$ \\
   & $0.93$ & $97.26$ \\
   Retrograde & $0.03$ & $99.91$ \\
 & $0.52$ & $99.70$ \\
   & $0.93$ & $99.22$ \\
\hline
\end{tabular}
\end{table}

\begin{figure*}
\centerline{\includegraphics[width=33pc]{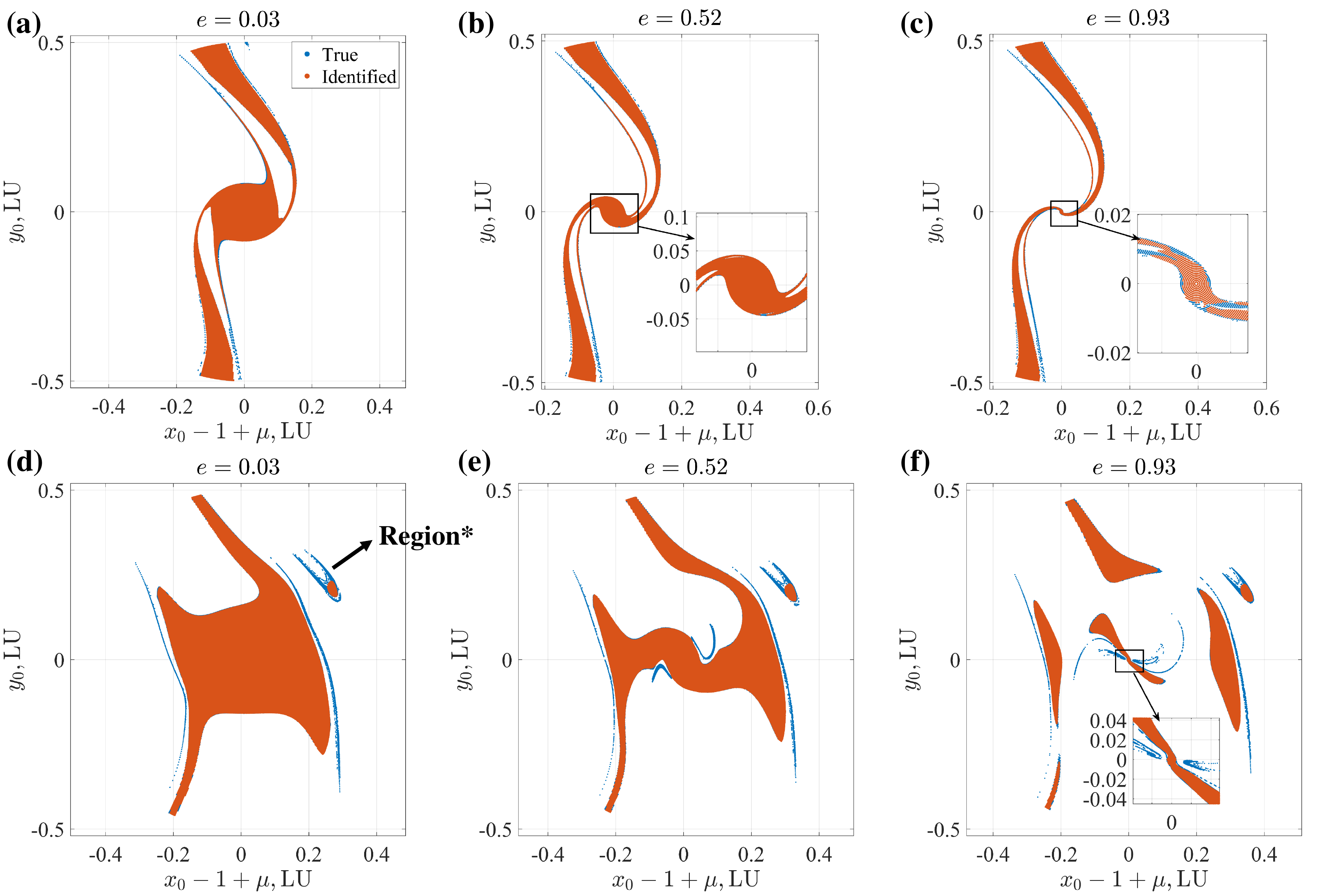}}
\caption{Identified WSB structures and corresponding true values. (a) Prograde, $e=0.03$; (b) Prograde, $e=0.52$; (c) Prograde, $e=0.93$; (d) Retrograde, $e=0.03$; (e) Retrograde, $e=0.52$; (f) Retrograde, $e=0.93$.}\label{fig_other_WSB}
\end{figure*}

This validation can be also considered as the application of the obtained models to the construction of WSB structures. Figure \ref{fig_other_WSB} presents the identified WSB structures and corresponding true values. It can be found that the incorrect identification is typically concentrated on the boundary, scatter-like regions. Moreover, for the retrograde case, the incorrect identification is also distribute in the $\text{Region}^*$, which is labeled in Fig. \ref{fig_other_WSB} (d). These results further strengthen the effectiveness of the trained DNN models and proposed DNN-based identification method.

\section{CONCLUSION}\label{sec5}
This paper is devoted to proposing a efficient and precise identification method of weak stability boundary (WSB) structures using deep neural network (DNN). The Earth-Moon planar circular restricted three-body problem is adopted to construct WSB structures and corresponding samples. The feature of the samples is analyzed before training the DNN models. It is shown that progarde and retrograde WSB structures exhibit different geometric configurations and dynamical properties. Based on this analysis, the samples are divided into two datasets: one for prograde case and the other for retrograde case. The DNN models to identify these two cases are trained respectively. The optimal hyperparameter combinations of the DNN models for these two cases are determined by evaluating the identification precision under different settings of the considered hyperparameters. Finally, the performance of the selected DNN models are further evaluated using the representative test datasets and the DNN models are applied to the construction of WSB structures. The precision when identifying the WSB structures corresponding to these test datasets is 97.26-99.91$\%$, strengthening the effectiveness of the obtained DNN models and proposed method. This work can provide further insight into the use of WSB structures in the construction of ballistic capture and low-energy transfer trajectories, establishing a novel link between multi-body dynamics and artificial intelligence.

\section*{ACKNOWLEDGMENT}

The authors acknowledge the suggestions on code improvement from Siyuan Guo from School of Astronautics, Beihang University.

\bibliographystyle{IEEEtran}
\bibliography{sample}

\begin{IEEEbiography}[{\includegraphics[width=1in,height=1.25in,clip,keepaspectratio]{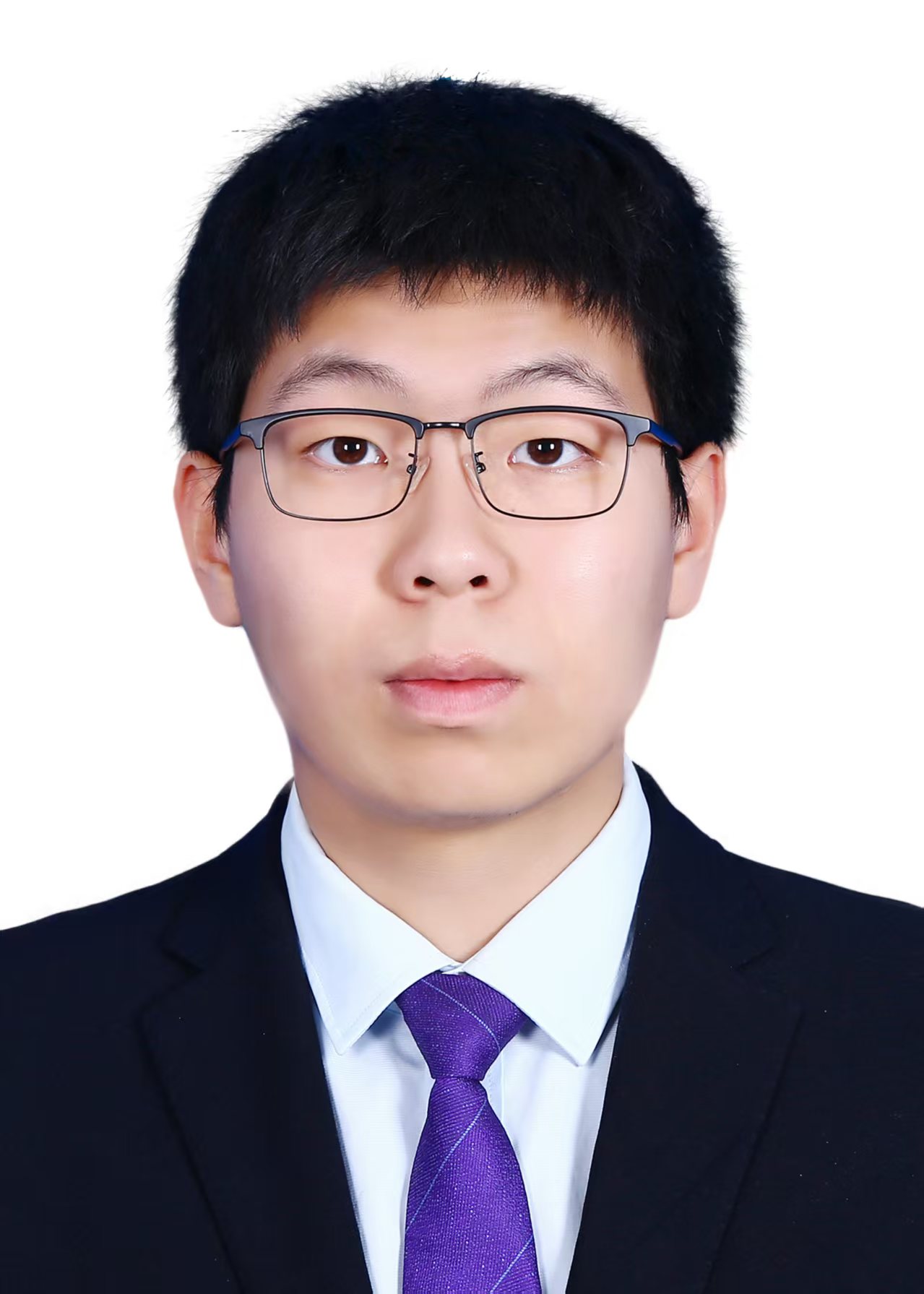}}]{Shuyue Fu} 
	received his B.S. degree in engineering in 2023 from Beihang University, Beijing, China, and
	he is currently working toward the Ph.D degree in astrodynamics and control with the School of Astronautics and Shen Yuan Honors College.
	His research interests include multi-body escape and capture dynamics, low-energy transfer, and orbital game.  
\end{IEEEbiography}

\begin{IEEEbiography}[{\includegraphics[width=1in,height=1.25in,clip,keepaspectratio]{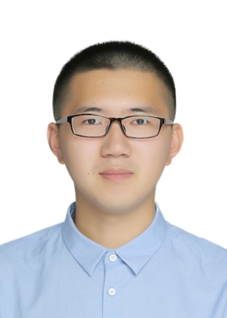}}]{Ziqi Xu} 
	received his B.S. degree in engineering in 2022 from Beihang University, Beijing, China, and
	he is currently working toward the Ph.D degree in astrodynamics and control with the School of Astronautics.
	His research interests include optimal control, the application of artificial intelligence technology to astrodynamics, trajectory optimization and hypersonic morphing vehicles.  
\end{IEEEbiography}

\begin{IEEEbiography}[{\includegraphics[width=1in,height=1.25in,clip,keepaspectratio]{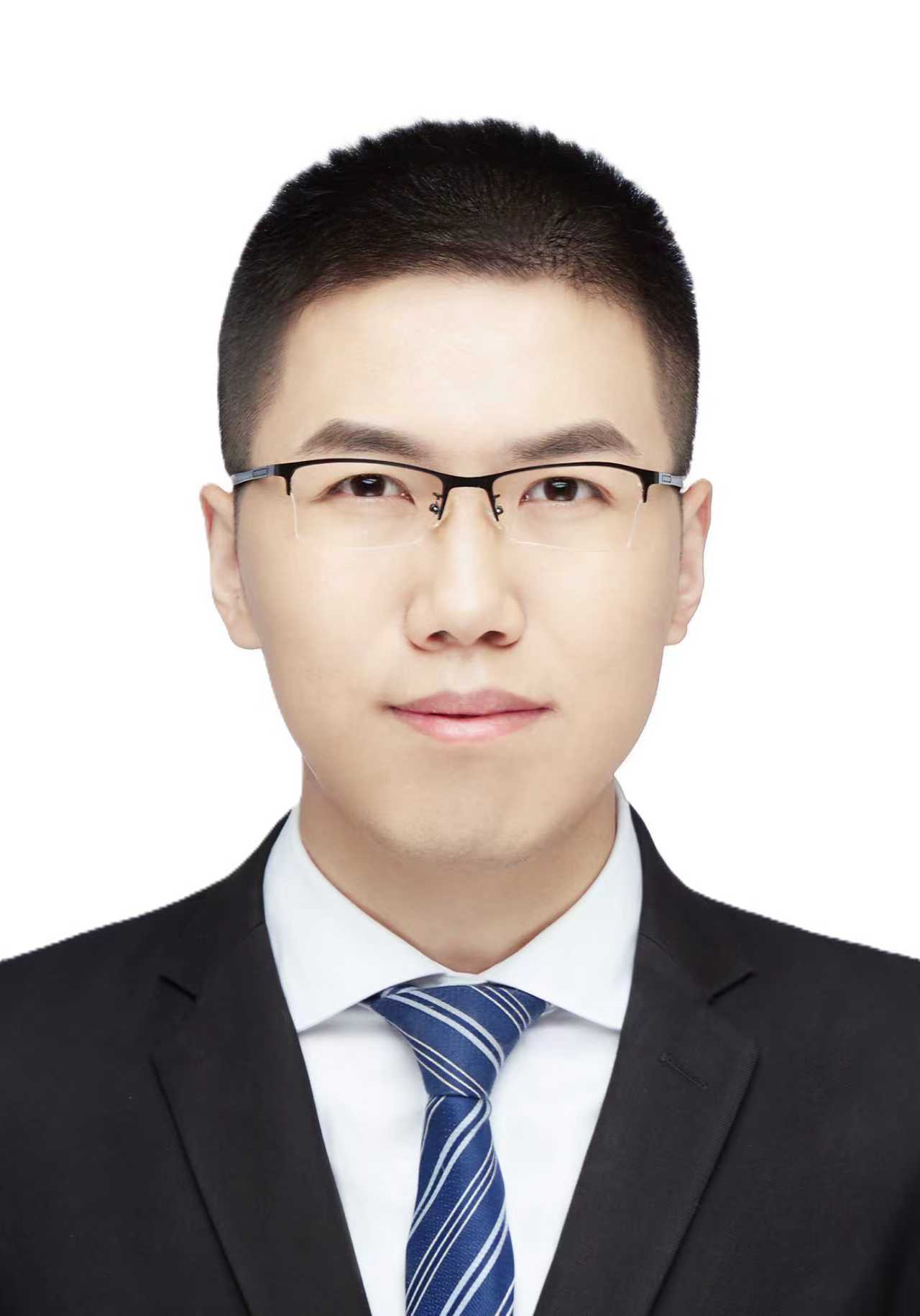}}]{Di Wu} received his Ph.D. degree, in 2022, from Tsinghua University. He is an associate professor at Beihang University, and his reasearch insterests include low-thrust trajectory optimization and mission design.
\end{IEEEbiography}

\begin{IEEEbiography}[{\includegraphics[width=1in,height=1.25in,clip,keepaspectratio]{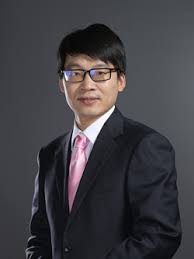}}]{Shengping Gong} received his B.S. degree in aerospace engineering from the National University of Defense Technology, Changsha, China, in 2004, and the Ph.D. degree in astrodynamics and control from Tsinghua University, Beijing, China, in 2008.

After spending a year as a Postdoctoral Researcher, he became an Assistant Researcher and then an Associate Professor with the School of Aerospace Engineering, Tsinghua University. In 2021, he was a Professor with the School of Astronautics, Beihang University. His research interests include the dynamics and control of spacecraft, trajectory optimization, solar sail, and celestial mechanics.
\end{IEEEbiography}

\end{document}